\documentclass{jpp}
\usepackage{graphicx}
\usepackage{natbib}

\usepackage[utf8]{inputenc}
\usepackage[T1]{fontenc}
\usepackage{amsmath}
\newcommand{\markup}[1]{{#1}}
\newcommand{\beq}{\begin{equation}}
\newcommand{\eeq}{\end{equation}}

\newcommand{\vu}{\boldsymbol{u}}
\newcommand{\vB}{\boldsymbol{B}}
\newcommand{\vb}{\boldsymbol{b}}

\newcommand{\vAy}{v_{\mathrm{A}y}}
\newcommand{\vA}{v_{\mathrm{A}}}

\newcommand{\vz}{\boldsymbol{z}}

\newcommand{\hg}{\hat{\Gamma}_0}
\newcommand{\zot}{\frac{Z}{\tau}}

\newcommand{\ddtof}[1]{\frac{d{#1}}{dt}}
\newcommand{\ddt}{\frac{d}{dt}}
\newcommand{\brak}[2]{\left\{{#1},{#2}\right\}}
\newcommand{\nbp}{\nabla_\perp}
\newcommand{\dtpe}{\delta T_{\parallel e}}
\newcommand{\vthe}{v_{\mathrm{th}e}}
\newcommand{\vthi}{v_{\mathrm{th}i}}
\newcommand{\din}{\delta_{in}}
\newcommand{\gammatr}{\gamma_{\mathrm{tr}}}
\DeclareMathOperator{\sech}{sech}

\shorttitle{Suppression of the collisionless tearing mode by shear flow}
\shortauthor{A. Mallet, S. Eriksson, M. Swisdak, J. Juno}

\title{Suppression of the collisionless tearing mode by flow shear: implications for reconnection onset in the Alfv\'enic solar wind}

\author{A. Mallet\aff{1}
  \corresp{\email{alfred.mallet@berkeley.edu}},
  S. Eriksson\aff{2},
 M. Swisdak\aff{3}, \and
 J. Juno\aff{4}
}

\affiliation{\aff{1}Space Sciences Laboratory, University of California, Berkeley CA 94720, USA
\aff{2}Laboratory for Atmospheric and Space Physics, University of Colorado, Boulder, CO 80303,USA
\aff{3}Institute for Research in Electronics and Applied Physics, University of Maryland, College Park, MD 20742, USA
\aff{4}Princeton Plasma Physics Laboratory, Princeton, NJ 08543, USA}
\begin{document}

\maketitle
\begin{abstract}
We analyse the collisionless tearing mode instability of a current sheet with a strong shear flow across the layer. The growth rate decreases with increasing shear flow, and is completely stabilized as the shear flow becomes Alfv\'enic. We also show that in the presence of strong flow shear, the tearing mode growth rate decreases with increasing background ion-to-electron temperature ratio, the opposite behaviour to the tearing mode without flow shear. We find that even a relatively small flow shear is enough to dramatically alter the scaling behaviour of the mode, because the growth rate is small compared to the shear flow across the ion scales (but large compared to shear flow across the electron scales). Our results may explain the relative absence of reconnection events in the near-Sun Alfv\'enic solar wind observed recently by NASA's Parker Solar Probe.
\end{abstract}

\section{Introduction}
Magnetic reconnection is a fundamental plasma physics process, involving a topological rearrangement of the magnetic field, and accompanied by the conversion of magnetic energy into bulk plasma flow and heat. Reconnection occurs in a wide range of contexts, for example solar flares \citep{yan2022}, at planetary magnetopauses \citep{paschmann2013}\markup{, in magnetic confinement fusion devices \citep{kadomtsev1975,zanini2020} and other laboratory plasmas \citep{ji2023}}, and in current sheets ubiquitous in many space and astrophysical environments \citep{eriksson2022,eriksson2024}.

Here, we are specifically motivated by observations of current sheets in the near-Sun solar wind, currently being explored \emph{in situ} for the first time by NASA's Parker Solar Probe (PSP) spacecraft. While such sheets are observed essentially ubiquitously \citep{vasko2022,lotekar2022}, only a small fraction actually reconnect \citep{eriksson2022}. 
\markup{In Alfv\'enic solar wind \citep{damicis2015,damicis2021}, the fluctuations are close to Alfv\'en waves propagating away from the Sun, with $\delta \vu \approx \pm\delta \vb$, where $\delta \boldsymbol{b}=\delta \vB/\sqrt{4\pi n_{i} m_i}$ is the magnetic field fluctuation in velocity units. An even more pronounced absence of reconnection events has been observed in this type of solar wind \citep{phan2020,fargette2023,eriksson2024}, suggesting that the quasi-Alfv\'enic velocity shear accompanying the magnetic shear across current sheets} suppresses the onset of reconnection, as observed in steady-state nonlinear reconnection \markup{simulations} \citep{cassak2011}. Here, we develop a \markup{linear} theory to describe the suppression of reconnection onset by such a flow shear.

Many numerical studies of reconnection begin with a kinetic scale current sheet: for example, the GEM reconnection challenge problem \citep{birn2001} has a current sheet with a width comparable to $d_i=c/\omega_{pi}$\markup{, where $\omega_{pi}=(4\pi n_i Z^2 e^2/m_i)^{1/2}$ is the ion plasma frequency, with $Z=q_i/e$}. With this setup, reconnection proceeds rapidly from the outset. However, in many of the natural systems mentioned above, reconnection occurs as part of a bursty, two-timescale process: first a long, quiescent phase in which magnetic energy builds up in thinning current sheets, followed by an extremely rapid disruption as reconnection occurs.

Recently, new models of reconnection onset that explain this property have been developed \citep{pucci2014,uzdensky2016,tolman2018}. On large scales, ideal dynamics lead to the progressive "thinning" of the current sheet width $a$, for example via the Chapman-Kendall collapse \citep{chapman1963}, on a characteristic (ideal) timescale $T(a)$. This can be compared with the growth rate of the tearing mode (the linear stage of reconnection), $\gamma(a)$: only when $\gamma(a) T(a) \gtrsim 1$
will reconnection onset occur: at which point, the ideal dynamics are disrupted and the current sheet is usually destroyed. This model has also been applied to turbulence, where $T$ can be identified as the nonlinear timescale $T_{\rm nl}$, whose scaling depends on the details of the turbulence \citep{msc_disruption,msc_cless,loureiroboldyrev,loureiroboldyrev_cless,comisso2018}.

To apply this model to the highly sheared current sheets observed by PSP, we need to understand the scalings of both $\gamma$ and $T$ in the presence of strong flow shear, neither of which are currently well understood. Here, we focus on the former problem, and develop a new analytic theory for the collisionless tearing mode with significant flow shear, $\delta u \sim \delta b = \delta B/\sqrt{4\pi n_{i} m_i}$, determining the growth rate $\gamma$ as a function of the physical parameters. We find that flow shear strongly suppresses the tearing instability, with the growth rate proportional to $1-\alpha^2$, where $\alpha = \delta u/\delta b$: as $\alpha\to1$, corresponding to Alfv\'enic flow shear, the growth rate of the tearing mode vanishes (for $\alpha>1$, we would have instead the ideal Kelvin-Helmholtz instability, whose growth rate is proportional to $\sqrt{\alpha^2-1}$: we will not comment further on this instability here). Finally, we discuss how our results might apply to the PSP observations.

\subsection{Tearing mode without shear flow}\label{sec:noflow}
To compare with the new results we will find in our calculation, we first present an overview of the collisionless tearing mode scalings without equilibrium flow shear\footnote{Specifically, the scalings at low $\beta$ with a strong guide field: these can be found from Eqs.~(\ref{eq:philin}--\ref{eq:gelin}) without the flow shear terms, i.e. with $\alpha=0$. A clear and concise derivation, along with references to previous works, is given by \citealt{zocco2011}.}. We assume that the current sheet normal is in the $\hat{\boldsymbol{x}}$ direction, with reconnecting field $\vAy=\delta b=\delta B / \sqrt{4\pi n_{0i} m_i}$ and the wavenumber of the tearing mode both in the $\hat{\boldsymbol{y}}$ direction, and a strong constant guide field $B_0\gg \delta B$ pointing in the $\hat{\vz}$ direction. The growth rate is given by
\begin{align}
    \frac{\gamma_0 a}{\vAy} \sim \begin{cases}
k\Delta' d_e \rho_s\sqrt{1+\tau/Z}, &\Delta'\delta_{in}\ll 1 \\
kd_e^{1/3}\rho_s^{2/3}\sqrt{1+\tau/Z}, &\Delta'\delta_{in}\gg 1,\label{eq:clessnoflow}
    \end{cases}
\end{align}
where $\rho_s = \sqrt{ZT_{0e}/m_i}/\Omega_i$ is the ion sound radius and $d_e=c/\omega_{pe}$ is the electron inertial length, with $Z=q_i/e$, $\tau=T_{0i}/T_{0e}$, $\omega_{pe}=(4\pi n_{0e}e^2/m_e)^{1/2}$ the electron plasma frequency and $\Omega_i=ZeB_0/m_i c$ the ion gyrofrequency \citep{zocco2011}. $\Delta'$ is a measure of the ideal discontinuity \markup{(see Sec.~\ref{sec:outer} for details)}, with $\Delta'>0$ required for instability, while $\din\ll a$ is the inner layer width over which the microphysics becomes important. \markup{The explicit form of $\din$ depends on the regime, but is not needed here. We refer the reader to \cite{zocco2011} for further details. $\Delta'$ depends on $k$ in a way that depends on the specific equilibrium profile \citep{boldyrev2018}:} for $ka\ll1$, we have $\Delta'a \propto (k a)^{-n}$ with $n=1$ for a Harris type equilibrium \markup{\citep{harris1962}} with $f(x/a) = \tanh(x/a)$ and $n=2$ for $f(x/a) = \sin(x/a)$. For $\Delta'\din\ll1$, the growth rate $\gamma_0$ decreases with $k$, while at $\Delta'\din\gg1$, $\gamma_0$ increases with $k$. The maximum growth rate and wavenumber at which it is attained may be found by equating \markup{the two expressions in Eq.~(\ref{eq:clessnoflow})},
\begin{equation}
    \frac{\gamma_{0 \mathrm{tr}}a}{\vAy} \sim \frac{\displaystyle d_e^{1/3+2/(3n)} \rho_s^{2/3+1/(3n)}\sqrt{1+\tau/Z}}{\displaystyle a^{1+1/n}}, \quad k_{0 \mathrm{tr}} a \sim \frac{\displaystyle d_e^{2/(3n)}\rho_s^{1/(3n)}}{\displaystyle a^{1/n}}.\label{eq:gmaxcless}
\end{equation}
For $n=1$ the growth rate in the $\Delta'\delta_{in}\ll1$ case does not depend on $k$, and so the growth rate is the same for all $k> k_{0\mathrm{tr}}$: for $n=2$ this is not the case. In the calculation that follows, we will find that these scalings are strongly affected by the presence of even relatively modest flow shear.

\section{Equations}
We will use the kinetic reduced electron heating model (KREHM) equations \citep{zocco2011}\markup{, in the collisionless limit}. These equations are derived from gyrokinetics (and thus assume low frequency fluctuations $\omega \ll \Omega_i$ and a strong guide field, $\delta B/B_0\ll1$) in the limit of \markup{small $\beta_e = 8\pi n_{0e}T_{0e}/B_0^2$}, and, assuming no variations in the $\hat{\vz}$ direction, may be written
\begin{align}
    \ddt(1-\hg)\Phi &=-\frac{1}{2}\rho_i^2\brak{\Psi}{\nbp^2\Psi},\label{eq:nezs}\\
    \ddt(\Psi - d_e^2 \nbp^2\Psi) &= \brak{\Psi}{\frac{Z}{\tau}(1-\hg)\Phi - \frac{c}{eB_0}\dtpe}\label{eq:Psizs}\\
    \ddtof{g_e} + \frac{v_\parallel}{v_A}\brak{\Psi}{g_e - \frac{\dtpe}{T_{0e}}F_{0e}} &= -\frac{1}{\Omega_i}\left(1-\frac{2v_\parallel^2}{\vthe^2}\right)F_{0e}\brak{\Psi}{\nbp^2\Psi}.\label{eq:gezs}
\end{align}
where the Poisson bracket \markup{is} $\{f,g\}=\hat{\boldsymbol{z}} \cdot(\nabla_\perp f \times \nabla_\perp g)$, 
\markup{and} the ion gyroradius is $\rho_i= \vthi/\Omega_i$, 
with $\vthi=\sqrt{2T_{0i}/m_i}$ the ion thermal speed
. $B_0$ is again the (strong) guide magnetic field in the out-of-plane $\hat{\vz}$ direction, $F_{0e}$ is the equilibrium Maxwellian electron distribution function, \markup{and} 
\begin{equation}
    \ddt = \frac{\partial}{\partial t} + \brak{\Phi}{\ldots}.
\end{equation}

The equations have been written in different variables to \markup{\cite{zocco2011}},
\beq
\Phi = \frac{c}{B_0}\phi, \quad \Psi = -\frac{A_\parallel}{\sqrt{4\pi m_i n_{0i}}},
\eeq
chosen to be notationally similar to reduced magnetohydrodynamics (RMHD): $\phi$ is the electric potential and $A_\parallel$ is the parallel magnetic vector potential, while the perpendicular (to $\hat{\vz})$ magnetic field (in velocity units) is $\vb_\perp = \hat{\vz}\times \nabla_\perp \Psi$ and the $\boldsymbol{E}\times\vB$ velocity is $\vu_{E\times B} = \hat{\vz}\times \nabla_\perp \Phi$. The operator $\hg$ is the inverse Fourier transform of
\beq
\Gamma_0(k_\perp^2\rho_i^2/2) = I_0(k_\perp^2\rho_i^2/2)e^{-k_\perp^2\rho_i^2/2},
\eeq
where $I_0$ is the modified Bessel function: at large and small scales, we have
\begin{equation}
    1-\hg \approx \begin{cases}
        -\frac{\rho_i^2}{2}\nbp^2, \quad &\rho_i^2\nbp^2\ll1\\
        1, \quad &\rho_i^2\nbp^2\gg1.
    \end{cases}
\end{equation}
\markup{In Eq.~(\ref{eq:nezs}) we have already substituted}
\beq
\frac{\delta n_e}{n_{0e}} = - \frac{2}{\rho_i^2\Omega_i}(1-\hg)\Phi = \frac{\delta n_i}{n_{0i}}
\eeq
\markup{for the density fluctuations.}
Equation (\ref{eq:gezs}) evolves the reduced parallel electron distribution function $g_e$, defined in terms of the perturbed parallel electron distribution function $\delta f_e$ as
\begin{equation}
    g_e = \delta f_e - \left[\frac{\delta n_e}{n_{0e}} + \frac{2v_\parallel u_{\parallel e}}{\vthe^2}\right]F_{0e},
\end{equation}
The parallel electron temperature fluctuation is given by the second moment of $g_e$,
\begin{equation}
    \frac{\dtpe}{T_{0e}} = \frac{1}{n_{0e}}\int d^3\boldsymbol{v} \frac{2v_\parallel^2}{\vthe^2}g_e.\label{eq:Tdef}
\end{equation}

KREHM is designed to be a set of equations appropriate for studying reconnection: the equations contain a rigorous treatment of electron heating via the electron kinetic equation (\ref{eq:gezs}), the dispersion at the ion scales ($\rho_i$ and $\rho_s = \rho_i\sqrt{Z/2\tau}$), important for achieving fast reconnection \citep{shay2001}, and also the flux-unfreezing at the electron inertial scale $d_e$. This model cannot be expected to apply to all the current sheets observed in the solar wind: in particular, the assumptions that $\beta\ll1$ and $\delta B/B_0\ll1$ are often not satisfied, \markup{nor do we attempt to model sheared parallel flow or density gradients across the current sheet}. However, it will allow us to make progress in understanding reconnection onset in the near-Sun solar wind.

\subsection{Equilibrium and linearized equations}
Our chosen equilibrium is
\begin{equation}
\Phi_0 = \alpha \Psi_0, \quad b_{0y} = \vAy f(x/a) = \partial_x \Psi_0,
\end{equation}
where we will be especially interested in the case where the shear flow in the $\hat{\boldsymbol{y}}$ direction is \markup{comparable to the reconnecting magnetic field}, $\alpha\sim1$. As in \markup{Sec.~\ref{sec:noflow}}, we have the current sheet normal in the $\hat{\boldsymbol{x}}$ direction and both the reconnecting field and wavenumber in the $\hat{\boldsymbol{y}}$ direction. We also assume that the equilibrium length-scale is large compared to ion and electron scales, $a\gg\rho_i\sim \rho_s\gg d_e$. \markup{We have also implicitly assumed that $u_{0y},b_{0y}\ll v_A$, so we are limited to guide field reconnection. Relaxing this assumption is not possible with the KREHM equations, which assume that there are no flows on the same level as the background (guide) magnetic field $B_0$}. Linearizing Eqs.~(\ref{eq:nezs}--\ref{eq:gezs}) and assuming fluctuations of the form
\begin{align}
\delta \Phi = \Phi(x) \exp(iky + \gamma t),\quad 
\delta \Psi = \Psi(x) \exp(iky + \gamma t),
\end{align}
and that $k^{-1}\gg \rho_i, d_e$, we obtain
\begin{align}
    &(\gamma +i\alpha k \vAy f)(1-\hg)\Phi + \frac{1}{2}i\alpha k \vAy \rho_i^2f''\Phi = - \frac{1}{2}ik\vAy f\rho_i^2 \left[\Psi'' - k^2\Psi - \frac{f''}{f}\Psi\right],\label{eq:philin}\\
    &(\gamma + i \alpha k \vAy f)(\Psi-d_e^2\Psi'')= ik\vAy f\left[\left(1+\zot(1-\hg)\right)\Phi - \frac{c}{eB_0}\dtpe\right],\label{eq:psilin}\\
    &(\gamma + i\alpha k \vAy f)g_e + ik\vAy f \frac{v_\parallel}{v_A}\left(g_e - \frac{\dtpe}{T_{0e}}F_{0e}\right)=\nonumber\\
    &\quad\quad\quad\quad\quad\quad\quad\quad\quad\quad\quad\quad-\frac{1}{\Omega_i}ik\vAy f\left(1-\frac{2v_\parallel^2}{\vthe^2}\right)F_{0e}\left[\Psi'' - k^2\Psi - \frac{f''}{f}\Psi\right].\label{eq:gelin}
\end{align}

We first solve the linearized kinetic equation (\ref{eq:gelin}) for $g_e$, integrating according to (\ref{eq:Tdef}) to find
\begin{equation}
    \frac{\dtpe}{T_{0 e}} = -\frac{2}{\Omega_i}\frac{\vA}{\vthe}\frac{Z(\zeta)+\zeta Z'(\zeta)}{Z'(\zeta)}\left(\Psi'' - k^2\Psi - \frac{f''}{f}\Psi\right),
\end{equation}
where 
\begin{equation}
    \zeta = \left[\frac{i\gamma}{k\vAy f} -\alpha\right]\frac{\vA}{\vthe}
\end{equation}
and $Z(\zeta)$ is the plasma dispersion function, with $Z'(\zeta)=-2(1+\zeta Z(\zeta))$. Using (\ref{eq:philin}),
\begin{equation}
    \frac{c}{eB_0}\dtpe = -(G-1)\zot (1-\hg)\Phi,\label{eq:Tfluc}
\end{equation}
where we have dropped a term since $\rho_s^2 f''/f\ll1$, and
\begin{equation}
    G=2\left(\zeta^2 -\frac{1}{Z'(\zeta)}\right).\label{eq:Gdef}
\end{equation}
For $\zeta\to\infty$, $G\to 3$ (adiabatic electrons), while as $\zeta\to0$, $G\to1$ (isothermal electrons). Inserting (\ref{eq:Tfluc}) into (\ref{eq:psilin}), we obtain
\begin{equation}
    (\gamma + i \alpha k \vAy f)(\Psi-d_e^2\Psi'')= ik\vAy f\left[\left(1+G\zot(1-\hg)\right)\Phi\right].\label{eq:psilin2}
\end{equation}
We will solve Eqs.~(\ref{eq:philin}) and (\ref{eq:psilin2}) in the "outer region" $x\sim a$ and in the "inner region" close to $x=0$. \markup{Compared to the resistive MHD case \citep{chen1989}, the main differences are, first, that we have the electron inertia term involving $d_e$ allowing reconnection instead of resistivity, and second, the terms involving $1-\hg$ which encode the ion-scale behaviour.} Since there are two microscales in the problem, $\rho_s$ (or $\rho_i$) and $d_e\ll \rho_s$, there will be nested ion and electron boundary layers. Because $\gamma$ is real, the real and imaginary parts of the eigenmodes will be even and odd respectively around $x=0$, and it will turn out that the imaginary part is small compared to the real part.

\section{Outer region}\label{sec:outer}
Here, $x\sim a \sim (f'/f)^{-1} \sim k^{-1} \gg \rho_s, d_e$. On these scales,
\beq
1-\hg \approx -\frac{1}{2}\rho_i^2 \nabla_\perp^2 \ll 1,
\eeq
so it may be neglected in (\ref{eq:psilin2}) (but not \markup{in} \ref{eq:philin}, where all the terms are at least this small). Assuming $\gamma \ll k\vAy$, we may also neglect the growth terms. (\ref{eq:psilin2}) becomes
\begin{equation}
    \Phi=\alpha\Psi,\label{eq:phiouter}
\end{equation}
and, inserting this into (\ref{eq:philin}), we obtain
\begin{equation}
    (1-\alpha^2)f [\Psi'' - k^2\Psi - (f''/f)\Psi]=0,\label{eq:outer}
\end{equation}
so that the outer solution for $\Psi$ is the same as in the MHD tearing mode. As $x\to 0$, $\partial_x \gg k$ and $f\approx x/a$, and of the outer equation all we are left with is $\Psi''=0$, whence the real and imaginary parts are
\begin{align}
\Psi_R &\to \Psi_{\infty R}(1+ \frac{1}{2}\Delta'|x|), \quad x\to 0,\nonumber\\
\Psi_I &\to \Psi_{\infty I}(\frac{1}{2}\Delta' x \pm 1), 
\quad x\to 0\label{eq:outersol}
\end{align}
defining $\Delta' = {[\Psi']_{-0}^{+0}}/{\Psi(0)}$, the discontinuity in the outer solution's magnetic field.
\section{Inner region}
In the inner region, of width $\din\ll a$, the ion and electron-scale effects become important. Here, $x\ll a$, $f\approx x/a$ and  $\partial^2/\partial x^2 \gg k^2, f''/f$.
Defining
\begin{equation}
    \delta = \frac{\gamma a}{k\vAy},\label{eq:delta}
\end{equation}
we obtain the inner region equations
\begin{align}
    \left(\delta + i\alpha x\right)(1-\hg)\Phi &= - \frac{1}{2}ix\rho_i^2\Psi'',\label{eq:phiinner}\\
    \left(\delta + i\alpha x\right)(\Psi-d_e^2\Psi'')&=ix\left[\Phi + G\zot(1-\hg)\Phi\right].\label{eq:psiinner}
\end{align}
We need to deal with the nonlocal operator $1-\hg$: substituting (\ref{eq:phiinner}) into (\ref{eq:psiinner}) we obtain
\begin{equation}
    \left(\delta + i\alpha x\right)(\Psi-d_e^2\Psi'')=ix\Phi + x^2 G\rho_s^2\frac{\Psi''}{\delta + i\alpha x},\label{eq:psiinner2}
\end{equation}
but we still need to calculate $1-\hg$ to solve (\ref{eq:phiinner}) itself. One simplification is to take cold ions: as $\rho_i^2\partial_x^2\to0$, $1-\hg\to-(1/2)\rho_i^2\partial_x^2$ and we recover a MHD-like version of (\ref{eq:phiinner}). 
Instead, \markup{following \cite{pegoraro1986} and more recently \cite{zocco2011}}, we incorporate the hot-ion response (non-rigorously) by using the Pad\'e approximant,
\begin{equation}
    1-\hg \approx \frac{-(1/2) \rho_i^2\partial_x^2}{1-(1/2) \rho_i^2\partial_x^2}.
\end{equation}
Then, (\ref{eq:phiinner}) becomes
\begin{equation}
\Phi'' = \frac{ix\Psi''}{\delta + i\alpha x} - \frac{1}{2}i\rho_i^2\left(\frac{x\Psi''}{\delta + i\alpha x}\right)''.\label{eq:phiinner2}
\end{equation}
We now rescale our equations by the inner lengthscale $\din$, to be determined later: writing
\begin{equation}
    \xi = \frac{x}{\din}, \quad \lambda = \frac{\delta}{\din}, \quad \epsilon = \frac{d_e}{\rho_s}\ll1\label{eq:lambda}
\end{equation}
and with $'$ now denoting differentiation by $\xi$, (\ref{eq:phiinner2}) and (\ref{eq:psiinner2}) become
\begin{align}
    \Phi'' = \frac{i\xi\Psi''}{\lambda + i\alpha \xi} - \frac{1}{2}i\frac{\rho_i^2}{\din^2}\left(\frac{\xi\Psi''}{\lambda + i\alpha \xi}\right)'',\label{eq:phinorm} \\
    (\lambda + i\alpha\xi)\Psi - i \xi \Phi = \frac{\rho_s^2}{\din^2}\left[G\xi^2 + (\lambda + i\alpha\xi)^2\epsilon^2\right]\frac{\Psi''}{\lambda + i\alpha\xi},\label{eq:psinorm}
\end{align}
where the argument of $G(\zeta)$ (see Eq.~\ref{eq:Gdef}) is
\begin{equation}
    \zeta = \frac{1}{\sqrt{2}}\epsilon\left(\frac{i\lambda}{\xi} - \alpha\right).
\end{equation}
The electron inertia term in (\ref{eq:psinorm}) becomes important when $\xi\sim\lambda\epsilon$ ($x\sim \delta \epsilon$): this is also the scale below which the electrons are no longer isothermal, i.e. when $\zeta\sim1$ so that $G(\zeta)$ starts to differ from $1$. We are free to choose $\din$ to be of the same order as the ion scale, $\din\sim\rho_s$, and anticipate $\lambda \sim \epsilon \ll 1$.
\subsection{Ion layer}
We first solve the equations on the ion scales, $\xi\sim 1$. Since $\lambda\ll\alpha\sim 1$, to lowest order the equations are
\begin{align}
    \Phi'' = \frac{1}{\alpha}\left(\Psi'' - \frac{1}{2}\frac{\rho_i^2}{\din^2}\Psi''''\right),\label{eq:ionphi}\\
    \alpha\Psi - \Phi = -\frac{1}{\alpha}\frac{\rho_s^2}{\din^2}\Psi''.\label{eq:ionpsi}
\end{align}
The solution that matches onto the outer layer solution (\ref{eq:outersol}) is
\begin{align}
    \Psi_{i} &= \Psi_\infty\left(1+\frac{1}{2}\Delta'\delta_{in}\xi\right) + C_i e^{-\xi},\\
    \Phi_i &= \alpha \Psi_\infty\left(1+\frac{1}{2}\Delta'\delta_{in}\xi\right) + \frac{1}{\alpha}C_i e^{-\xi}\left[1-\frac{1-\alpha^2}{1+Z/\tau}\right],
\end{align}
where we have set
\begin{equation}
    \din = \frac{\rho_s\sqrt{1+\tau/Z}}{\sqrt{1-\alpha^2}}.\label{eq:din}
\end{equation}
\markup{It is worth noticing that the solution for $\Phi_i$ is very different from the tearing mode without shear as $\xi\to\infty$, to match the outer solution (\ref{eq:outersol}), which is also very different from its equivalent with $\alpha=0$, since the flow shear dominates over the growth rate on the large equilibrium scales. Moreover, it is clear that Eq.~(\ref{eq:ionphi}) depends strongly on the ratio of $\rho_i$ to $\rho_s$, i.e. on $\tau$. These two facts will conspire to change the $\tau$-dependence of the growth rate from the case with no shear flow (Eq.~\ref{eq:clessnoflow}).} As $\xi\to 0$, we have
\begin{align}
    \Psi_i &\to \Psi_\infty + C_i + \left(\frac{1}{2}\Delta'\delta_{in}\Psi_\infty-C_i\right)\xi,\label{eq:psiionlim}\\
    \Phi_i &\to \alpha \Psi_\infty +\frac{1}{\alpha}C_i \left[1-\frac{1-\alpha^2}{1+Z/\tau}\right] + \left(\frac{1}{2}\alpha\Delta'\delta_{in}\Psi_\infty-\frac{1}{\alpha}C_i\left[1-\frac{1-\alpha^2}{1+Z/\tau}\right]\right)\xi.\label{eq:phiionlim}
\end{align}
\subsection{Electron layer}
We rescale the equations again to find the solution on electron scales, defining
\begin{equation}
    y = \frac{\xi}{\lambda\epsilon} = \frac{x}{\delta \epsilon}.
\end{equation}
The equations are
\begin{align}
    \Phi'' &= \frac{i\epsilon y \Psi''}{1+i\alpha\epsilon y} - \frac{1}{\lambda^2\epsilon^2}\frac{1-\alpha^2}{1+Z/\tau} \left(\frac{i\epsilon y \Psi''}{1+i\alpha\epsilon y}\right)'',\label{eq:phielectron1}\\
    \lambda^2(1+i\alpha\epsilon y)\Psi -i\lambda^2\epsilon y \Phi &= \frac{1-\alpha^2}{1+\tau/Z}\left[(1+i\alpha\epsilon y)^2+Gy^2\right]\frac{\Psi''}{1+i\alpha\epsilon y}.\label{eq:psielectron1}
\end{align}
On these scales, the shear terms are small compared to the growth terms. Dividing (\ref{eq:psielectron1}) by $y$, differentiating twice and substituting for $\Phi''$ using (\ref{eq:phielectron1}), we obtain an equation for $\Psi$,
\begin{equation}
    \lambda^2\left[\left({1}/{y}+i\alpha\epsilon\right)\Psi\right]'' + \lambda^2\epsilon^2\frac{\Psi''}{1/y+i\alpha\epsilon} = \frac{1-\alpha^2}{1+\tau/Z}\left[\left(G+\frac{\tau}{Z}+\frac{1}{y^2}\left(1+i\alpha\epsilon y\right)^2\right)\frac{\Psi''}{1/y+i\alpha\epsilon}\right]''.
\end{equation}
Anticipating $\lambda\sim\epsilon$, we can now solve order-by-order. 
\markup{At lowest order, using (\ref{eq:psielectron1}), we have $\Psi_0''=0$, and so from (\ref{eq:phielectron1}) we also have $\Phi_0''=0$. To match the solution at larger scales, the lowest-order solution must be real and even, and we have}
\begin{equation}
    \Psi_0=\Psi_{0e}, \quad \Phi_0 = \Phi_{0e},
\end{equation}
both constants.
At first order, $\Psi_1''=0$, and we take $\Psi_1=0$: we can absorb any even constant piece into the zeroth-order solution, and since the solution is even and real to lowest order overall, any term linear in $y$ must appear at order $\lambda \epsilon$ or higher. At second order, we obtain
\begin{equation}
    \Psi_2'' = \frac{\lambda^2(1+\tau/Z)}{1-\alpha^2}\Psi_{0e} \frac{1}{(G+\tau/Z)y^2+1},
\end{equation}
so that 
\begin{equation}
    \Psi_2 = \frac{\lambda^2\sqrt{1+\tau/Z}}{1-\alpha^2}\Psi_{0e} \int_0^y dz \int_0^z\frac{du\sqrt{1+\tau/Z}}{(G+\tau/Z)u^2+1}.
\end{equation}
As $y\to\infty$, we then have (in terms of $\xi$),
\begin{equation}
    \Psi_e\to \Psi_{0e}\left[1 + \frac{\lambda\sqrt{1+\tau/Z}}{\epsilon(1-\alpha^2)}I_G\xi - \frac{\lambda^2\sqrt{1+\tau/Z}}{1-\alpha^2}\log(\xi)\right],\label{eq:psielectronlim}
\end{equation}
where
\begin{equation}
    I_G = \int_0^\infty \frac{dy\sqrt{1+\tau/Z}}{(G+\tau/Z)y^2+1}.
\end{equation}
If we had assumed isothermal electrons ($G=1$), $I_G=\pi/2$. The solution for $\Phi$ can be found by integrating (\ref{eq:phielectron1}) twice:
\begin{equation}
    \Phi = \Phi_{0e}- \frac{1}{\lambda^2\epsilon^2}\frac{1-\alpha^2}{1+Z/\tau} \left(\frac{i\epsilon y \Psi''}{1+i\alpha\epsilon y}\right)+i\epsilon\int_0^y dz \int_0^z du \frac{ u \Psi''}{1+i\alpha\epsilon u},\label{eq:phielectron2}
\end{equation}
The second term on the RHS, resulting from the small-scale asymptotic of $1-\hg$, is dominant for $y\sim1$, but always decays for $y\gg1$, since $\Psi''$ (e.g. $\Psi_2''$) is strongly peaked on the electron scales. The final term on the RHS does produce terms in the solution for $\Phi$ that do not decay as $y\to\infty$: however, note that they are still a factor $\epsilon$ smaller than the corresponding terms in $\Psi$. Inserting $\Psi_2''$ into (\ref{eq:phielectron2}), the lowest-order piece is, as $y\to\infty$ and in terms of $\xi$,
\begin{equation}
    \Phi_e \to \Phi_{0e}+i\epsilon\int_0^y dz \int_0^z du { u \Psi_2''} \to \Phi_{0e}+\frac{i\lambda(1+\tau/Z)}{1-\alpha^2}\Psi_{0e} \left[\xi\log(\xi)-\xi + \text{const.}\right] + \frac{i\lambda\Psi_{0e}}{1+Z/\tau}\frac{1}{\xi},\label{eq:phielectronlim}
\end{equation}
so the non-constant piece of $\Phi_e$ is $O(\lambda)$ for $\xi\sim1$. 

Finally, if we were to solve to third order for $\Psi$, we could determine the odd part of the eigenmode in terms of the even one, and thus determine the small imaginary part of the solution: because this appears at third order, the linear term that appears in the asymptotic solution as $y\to\infty$ is $\sim \lambda \xi$, showing that the imaginary part of the eigenmode is a factor of around $\lambda$ smaller than the real part. However, this is not necessary to obtain the growth rate.

\section{Matching and dispersion relation}
We can now match the ion and electron solutions. We match the constant and linear terms from $\xi\to0$ from the ion solution (\ref{eq:psiionlim} and \ref{eq:phiionlim}) and $y\to\infty$ in the electron solution (\ref{eq:psielectronlim} and \ref{eq:phielectronlim}), assuming $\Delta'\din\sim 1$. From the constant terms of $\Psi$, we obtain
\begin{equation}
    \Psi_{0e}= \Psi_\infty + C_i.\label{eq:psi0e}
\end{equation}
As we found above, at lowest order the leading non-constant terms of $\Phi_e$ are $O(\lambda)$, and so to lowest order, the linear term in (\ref{eq:phiionlim}) must be zero,
\begin{equation}
    \frac{1}{2}\alpha\Delta'\delta_{in}\Psi_\infty-\frac{1}{\alpha}C_i\left[1-\frac{1-\alpha^2}{1+Z/\tau}\right]=0,
\end{equation}
whence
\begin{equation}
    C_i = \frac{\frac{1}{2}\alpha^2\Delta'\delta_{in}\Psi_\infty(1+Z/\tau)}{Z/\tau + \alpha^2}.\label{eq:Ci}
\end{equation}
\markup{The $O(\lambda)$ terms of $\Phi_e$ in (\ref{eq:phielectronlim}) can in principle be matched with the next-order solution to $\Phi_i$ (which we have not calculated), providing a small correction to our results, but we do not attempt this here.} Finally, matching the linear terms for $\Psi_i$ as $\xi\to0$ (\ref{eq:psiionlim}) and $\Psi_e$ as $y\to\infty$ (\ref{eq:psielectronlim}),
\begin{equation}
    \frac{1}{2}\Delta'\delta_{in}\Psi_\infty-C_i = \frac{\lambda\sqrt{1+\tau/Z}}{\epsilon(1-\alpha^2)}I_G\Psi_{0e}.
\end{equation}
Inserting (\ref{eq:psi0e}) and (\ref{eq:Ci}), we obtain the dispersion relation
\begin{equation}
    \lambda = \frac{\frac{1}{2}\Delta'\delta_{in}\epsilon(1-\alpha^2)^2Z/\tau}{Z/\tau + \alpha^2 +\frac{1}{2}\alpha^2\Delta'\delta_{in}(1+Z/\tau)}\frac{1}{I_G\sqrt{1+\tau/Z}}.
\end{equation}

\subsection{Growth rate}
Using the definitions of $\lambda$ (\ref{eq:lambda}) and $\delta$ (\ref{eq:delta}), the growth rate is
\begin{equation}
    \frac{\gamma a}{\vAy} = \frac{k\Delta'\din^2\epsilon(1-\alpha^2)^2}{2 I_G \sqrt{1+\tau/Z}(1+\alpha^2\tau/Z+\frac{1}{2}\alpha^2\Delta'\din(1+\tau/Z))}
\end{equation}
Inserting $\din$ (\ref{eq:din}), we find that,
\begin{equation}
    \frac{\gamma a}{\vAy}\sim \begin{cases}
        k\Delta' \rho_s d_e \frac{\displaystyle(1-\alpha^2)\sqrt{1+\tau/Z})}{\displaystyle 2I_G(1 + \alpha^2\tau/Z)}, &\Delta'\din \ll 1\\
        k d_e \frac{\displaystyle(1-\alpha^2)^{3/2}}{\displaystyle\alpha^2 I_G(1+\tau/Z)}, &\Delta'\din \gg 1.
    \end{cases}\label{eq:cless_growthpred}
\end{equation}
The growth rate for $\Delta'\din\ll1$ has the same scaling with $d_e$ and $\rho_s$ as the no-flow scaling (\ref{eq:clessnoflow}), but for $\Delta'\din\gg1$, we have obtained the surprising result that the growth rate is independent of $\rho_s$: in the no-flow case, we had $\gamma_{0} \propto \rho_s^{2/3}d_e^{1/3}$. Note that the growth rate for $\Delta'\din\gg1$ does not match smoothly onto the no-flow scaling (\ref{eq:clessnoflow}): this is because the expansion in the ion region is invalidated for $\alpha \ll \lambda$ (see Sec.~\ref{sec:validity} below).

The transition between these scalings occurs where they match. Inserting $\Delta'a \propto 1/(ka)^n$ (see Sec.~\ref{sec:noflow}), the transition occurs roughly at 
\begin{equation}
    k_{\mathrm{tr}} a \sim (\rho_s/a)^{1/n} \frac{\alpha^{2/n}}{(1-\alpha^2)^{1/2n}}\left(\frac{(1+\tau/Z)^{3/2}}{1+\alpha^2\tau/Z}\right)^{1/n}\label{eq:kmax}
\end{equation}
for which the growth rate is of order
\begin{equation}
    \frac{\gammatr a}{\vAy} \sim \frac{\rho_s^{1/n}d_e}{a^{1+1/n}}\frac{(1-\alpha^2)^{3/2-1/2n}}{\alpha^{2-2/n}}\frac{(1+\tau/Z)^{3/2n-1}}{(1+\alpha^2\tau/Z)^{1/n}}.\label{eq:gmax}
\end{equation}
For $n=2$, the growth rate increases with $k$ for $\Delta'\din \gg1$ and decreases with $k$ for $\Delta'\din\ll1$, so that $\gammatr$ is also the maximum growth rate, occuring uniquely at $k_{\mathrm{tr}}$. For the $n=1$ case, $\gamma$ is independent of $k$ for $\Delta'\din\ll1$, and equal to $\gammatr$ for all $k>k_{\mathrm{tr}}$. 

This differs in several ways from the growth rate for $\alpha=0$ (Eq.~\ref{eq:clessnoflow}). First, the growth rate scales strongly with $1-\alpha^2$: as $\alpha\to1$, the growth rate vanishes, as is the case for the resistive MHD tearing mode with shear \markup{\citep{hofman1975,chen1989,boldyrev2018,shi2021b}}, where $\gamma_{\rm MHD}\propto (1-\alpha^2)^{1/2}$: thus the growth rate of the tearing mode is more strongly suppressed in a collisionless plasma than in resistive MHD.\footnote{For $\alpha^2>1$, we would have instead the Kelvin-Helmholtz instability: since $a\gg\rho_i$ this is essentially the same as the MHD case \citep{miura1982}.} Second, the growth rate depends differently on $\tau$. For $\tau\ll1$, the growth rate becomes independent of $\tau$, similarly to the tearing mode without flow shear. In contrast, for $\tau\gg1$, the maximum growth rate is proportional to $\tau^{(1/2n)-1}$; the opposite dependency to the case without flow shear, for which $\gamma_{0\mathrm{tr}}\propto \tau^{1/2}$ for $\tau\gg1$. Finally, in general the growth rate also depends differently on the scales $\rho_s$ and $d_e$ than in the no-flow case: without shear flow, $\gamma_{0\mathrm{tr}} \propto d_e^{1/3+2/3n}\rho_s^{2/3+1/3n}$, but with shear flow, $\gammatr\propto d_e \rho_s^{1/n}$. For $n=1$, these scalings coincide, but for $n=2$, $\gammatr/\gamma_{0\mathrm{tr}}\propto \epsilon^{1/3}$, where $\epsilon=d_e/\rho_s\ll1$.

\subsection{Validity and the small shear flow limit}\label{sec:validity}
For the expansion in the ion region to be valid, we must have
$\lambda \ll \alpha$.
For $\Delta'\din \ll1$, this is easily satisfied, and indeed, the growth rate in this case smoothly joins onto the no-flow growth rate (\ref{eq:clessnoflow}) as $\alpha\to0$. For $\Delta'\din\gg1$, the inequality is
\begin{equation}
    \frac{\epsilon(1-\alpha^2)^2}{\alpha^2 (1+\tau/Z)^{3/2}} \ll \alpha,
\end{equation}
or $\alpha$ above a critical $\alpha_c$ defined by
\begin{equation}
    \frac{\alpha_c}{(1-\alpha_c^2)^{2/3}} = \epsilon^{1/3}(1+\tau/Z)^{-1/2}.
\end{equation}
For $\epsilon\ll1$, $\alpha_c \sim \epsilon^{1/3}$. \markup{Taking $\beta_e\approx 0.1$ in the low-$\beta$ solar wind and corona, $\epsilon \approx0.1$ and $\alpha_c \approx 0.5$.
Thus, even a modest shear flow (far enough from $\alpha=1$ that $1-\alpha^2$ is not small) can affect the growth rate scalings with $d_e$ and $\rho_s$ for $\Delta'\din\gg1$.} This is because the growth rate of the mode is slow compared to the shearing rate on the ion scales.

\section{Numerical tests}
We use an eigenvalue code to solve (\ref{eq:philin}--\ref{eq:psilin}) assuming cold ions and isothermal electrons (and thus we do not need to solve Eq.~\ref{eq:gelin}): these assumptions (while unjustified physically for the case of the solar wind) make the equations much simpler to solve numerically. We use the profile \citep{loureiro2005}
 \begin{align}
     f(x/a) &= -2\tanh(x/a)\sech^2(x/a).
 \end{align}
 For $ka\ll1$, $\Delta'a \sim 15/(ka)^2$; i.e. $n=2$. For this profile, the maximum growth rate is attained uniquely at the transitional wavenumber, rather than for all $k>k_{\mathrm{tr}}$ as would be the case for the more usual $f(x/a)=\tanh(x/a)$ profile, for which $n=1$.
 \begin{figure}
     \centering
     \includegraphics[width=0.5\linewidth]{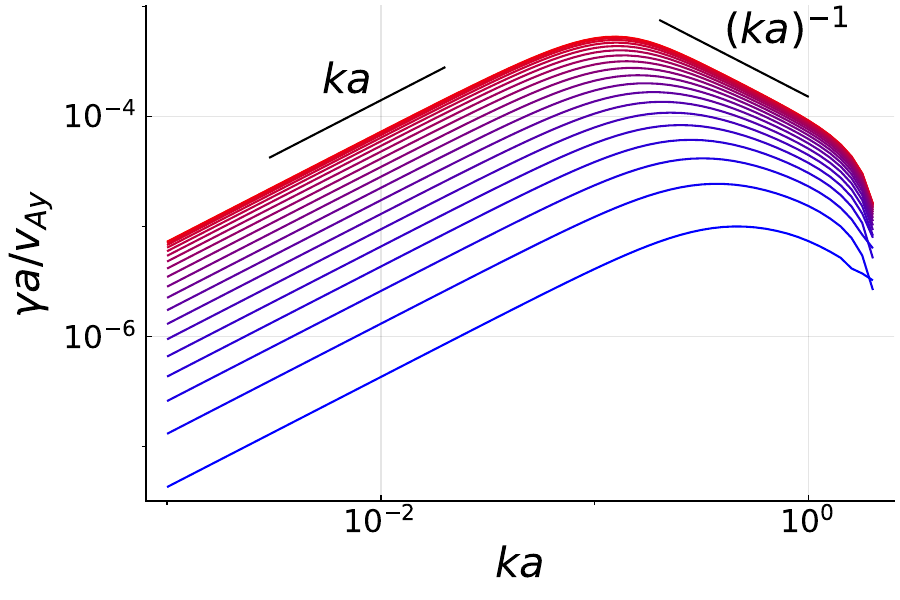}
     \caption{Growth rate as a function of $ka$ for $\alpha=0,0.05,\ldots,0.95$, red to blue lines. We set $\rho=0.01a$ and $d_e=0.001a$.}
     \label{fig:cless_gammasvsk_alphas}
 \end{figure}

Growth rates as a function of $k$ for different $\alpha$ are shown in Figure \ref{fig:cless_gammasvsk_alphas}, showing the expected scalings with $k$ for both small and large $k$ (large and small $\Delta'$). Because the transition moves to larger $k$ with increasing shear flow, the interval in $k$ over which the small-$\Delta'$ scalings are relevant gets narrower as $\alpha$ increases towards $1$.

\begin{figure}
    \centering
     \begin{minipage}{\linewidth}
     \includegraphics[width=0.45\linewidth]{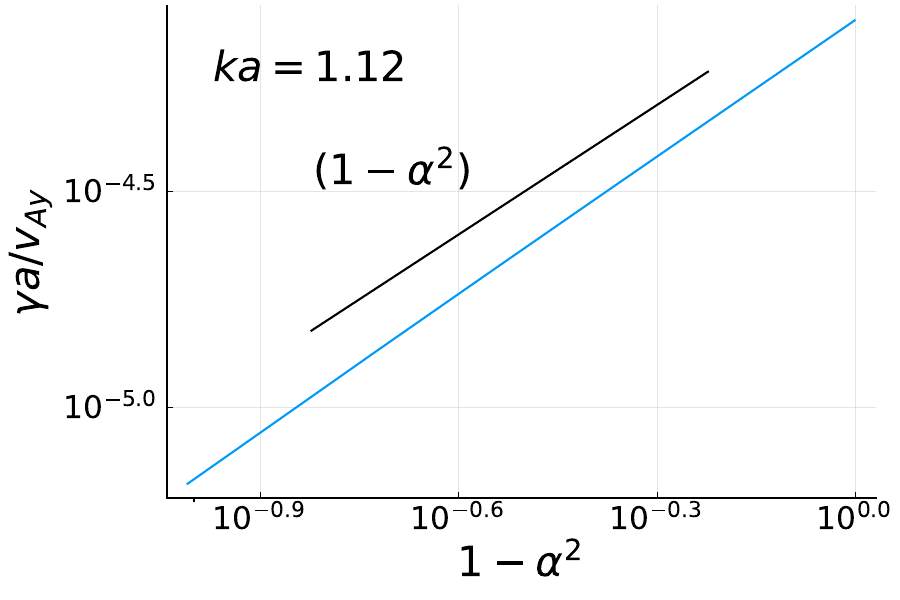}  \includegraphics[width=0.45\linewidth]{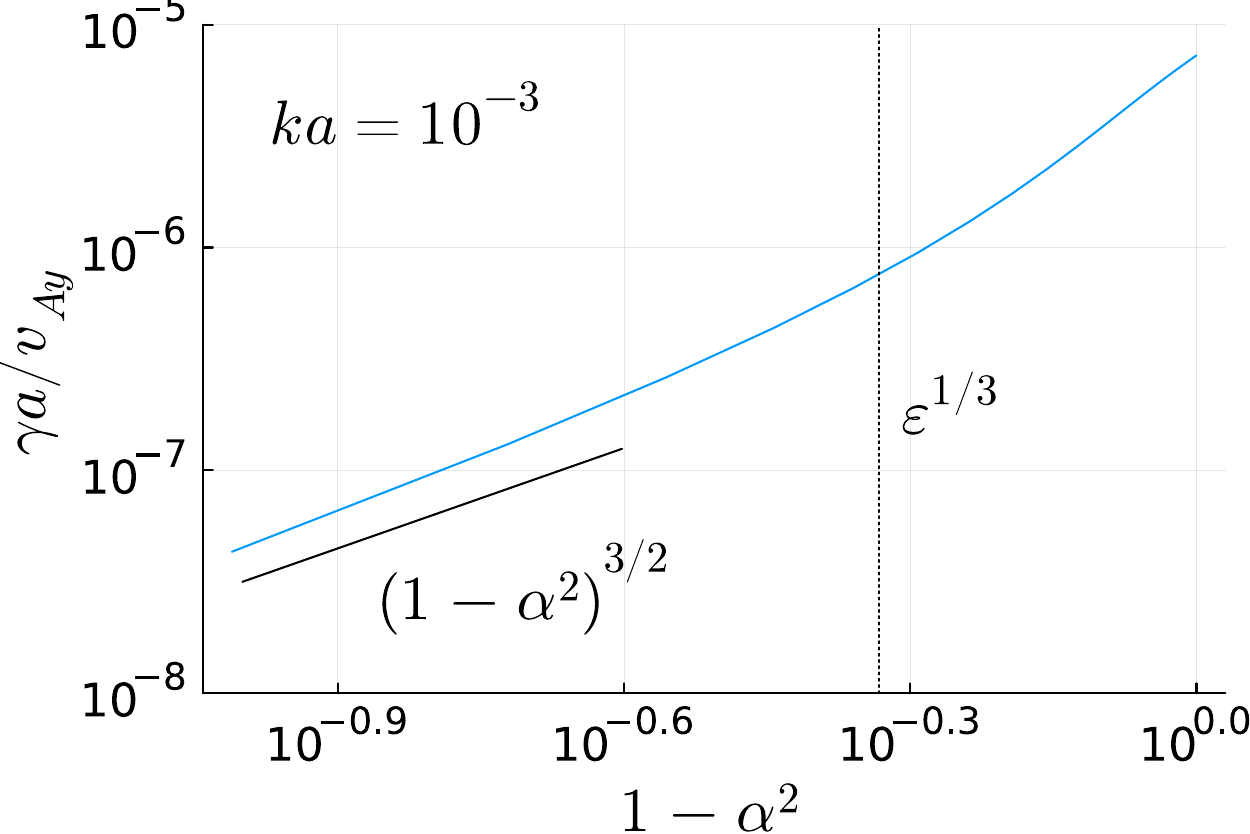} \\
    \includegraphics[width=0.45\linewidth]{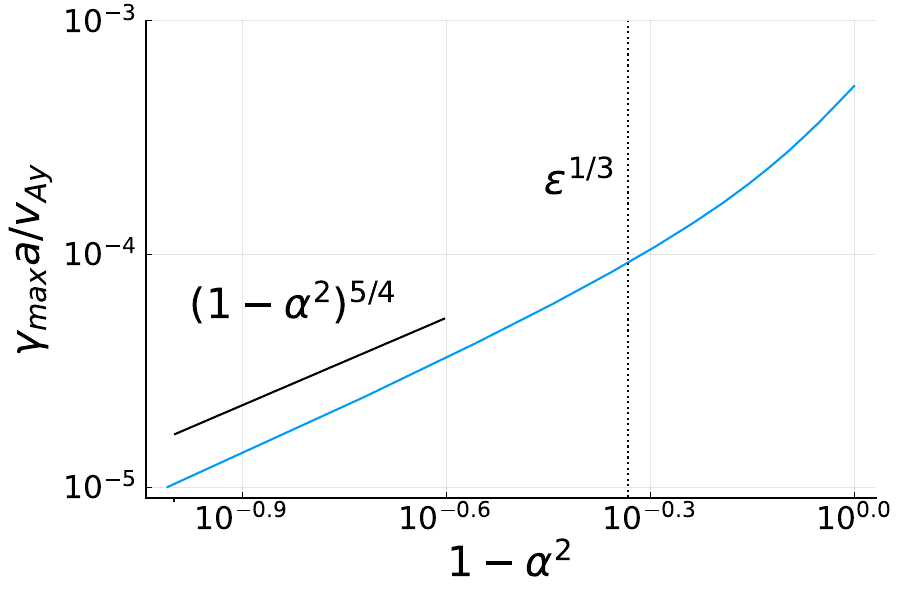} \includegraphics[width=0.45\linewidth]{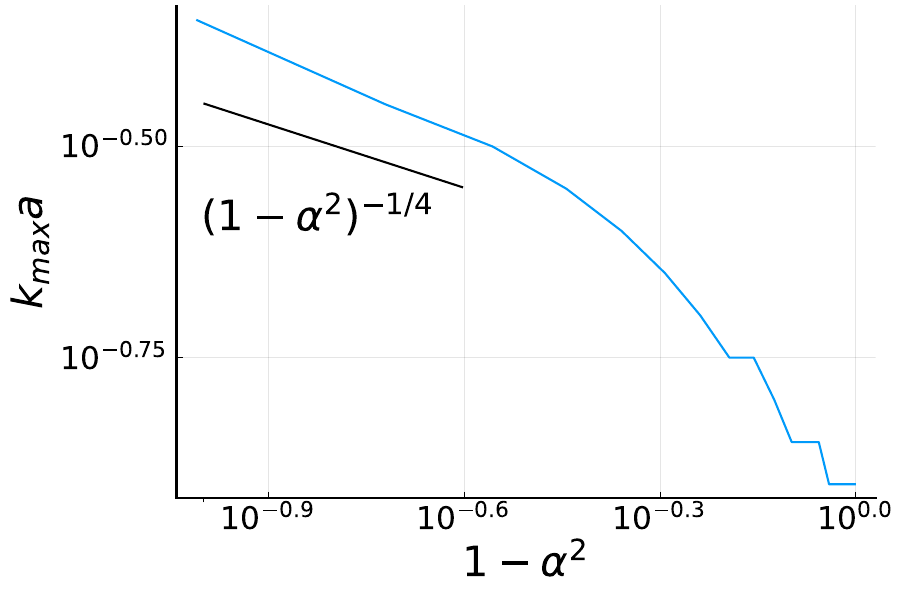}
     \end{minipage}
     \caption{Top left: $\gamma$ for $ka = 1.12$, at which wavenumber $\Delta'\din\ll1$. Top right: $\gamma$ for $ka=10^{-3}$ ($\Delta'\din\gg1$). Bottom left: maximum growth rate $\gamma_{max}$. Bottom right: the wavenumber $k_{\mathrm{tr}}$ at which $\gamma_{max}$ is attained. The vertical dashed lines on the top right and bottom left panels mark $\alpha=\epsilon^{1/3}$.}
     \label{fig:cless_alphascalings}
 \end{figure}

We plot growth rates at fixed $k$ as a function of $1-\alpha^2$ in Figure \ref{fig:cless_alphascalings}. For both small and large $\Delta'$ (top left and right panels respectively), the scalings agree with (\ref{eq:cless_growthpred}). In the top right panel, we have also marked on the x-axis the position at which $\alpha=\epsilon^{1/3}$: for smaller $\alpha$ (larger $1-\alpha^2$), we do not expect our scaling to apply - and indeed, the behaviour changes at around this point. The maximum/transitional growth rate (\ref{eq:gmax}) and corresponding transitional wavenumber (\ref{eq:kmax})  are shown in the bottom panels, and also agree quite well with the predicted scalings, shown as black lines.

\begin{figure}
    \centering
    \begin{minipage}{\linewidth}
    \includegraphics[width=0.45\linewidth]{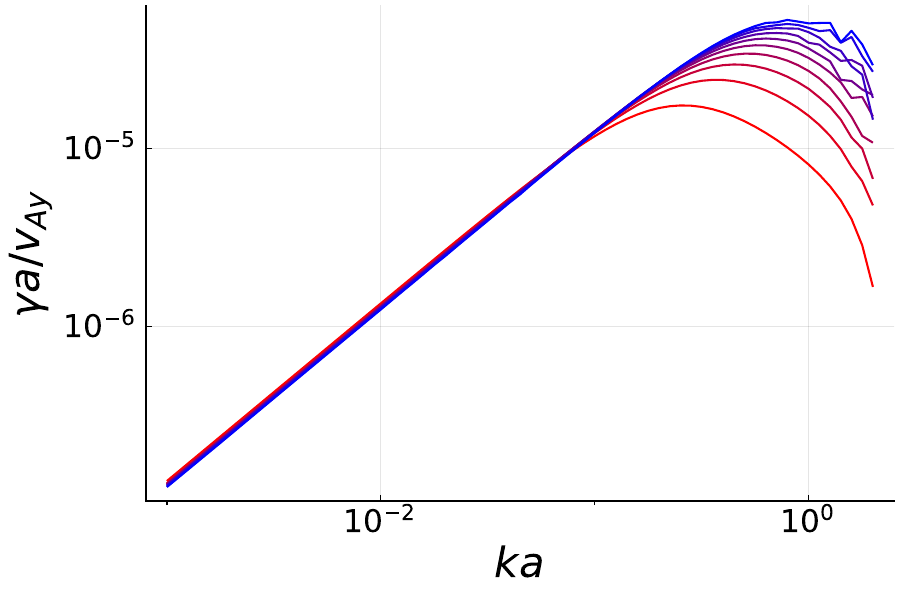}  \includegraphics[width=0.45\linewidth]{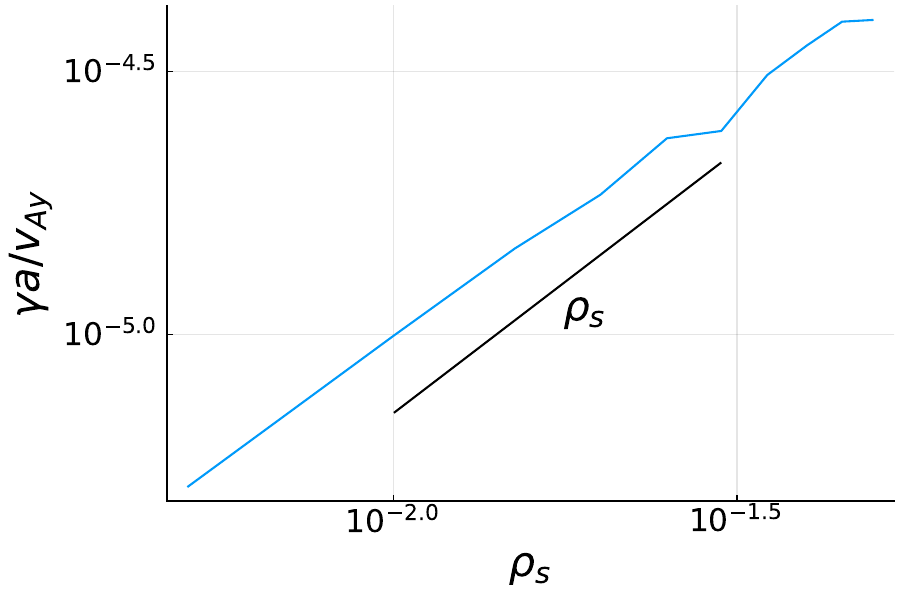} \\
    \includegraphics[width=0.45\linewidth]{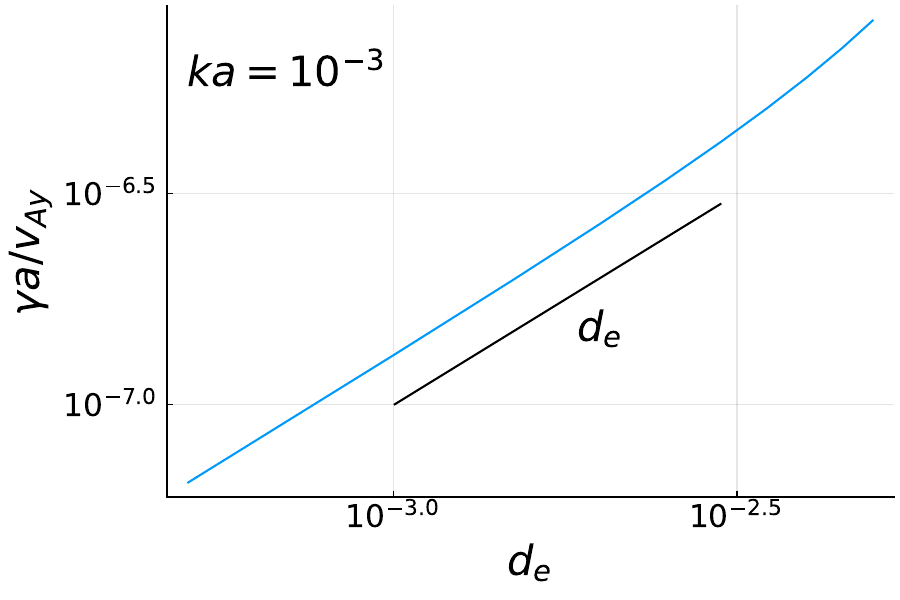} \includegraphics[width=0.45\linewidth]{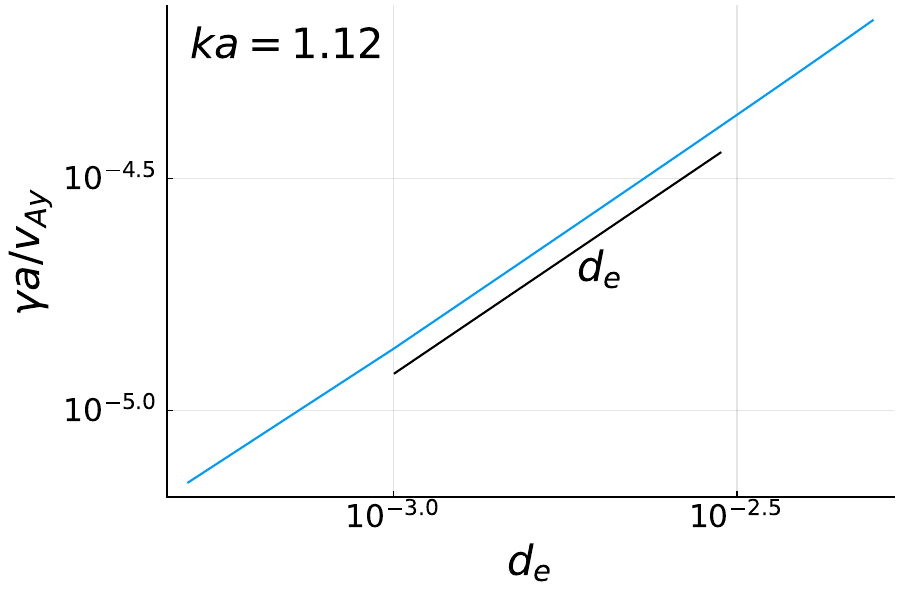}
    \end{minipage}
    \caption{In all panels $\alpha=0.9$. Top left: growth rate vs $k$, different lines correspond to different $\rho_s$. For large $\Delta'\din$ (smaller wavenumbers), the growth rate is independent of $\rho_s$. Top right: growth rate as a function of $\rho_s$ for $ka=1.12$, i.e. for $\Delta'\din\ll1$. Bottom left: growth rate as a function of $d_e$ for $ka=10^{-3}$, i.e. $\Delta'\din\gg1$. Bottom right: growth rate as a function of $d_e$ for $ka=1.12$.}
    \label{fig:cless_rhodescalings}
\end{figure}

We also check the dependence of the growth rate on both $\rho_s$ and $d_e$. As a reminder, for $\alpha=0$ the dependencies are given by (\ref{eq:clessnoflow}) with $\tau=0$: our code reproduces these scalings (not shown). In Figure \ref{fig:cless_rhodescalings} we plot the scalings for $\alpha=0.9$. For large $\Delta'$, the growth rate depends linearly on $d_e$, and does not depend on $\rho_s$: both as predicted (\ref{eq:cless_growthpred}), and very different to the $\alpha=0$ case. For small $\Delta'$, at a wavenumber of $ka=1.12$, the growth rate depends on the product $d_e \rho_s$, as in the no-flow case, also in agreement with our predictions.

We can also check the behaviour of the width of the perturbed current profile: this corresponds to the scale at which the electron terms become important, $x\sim\delta \epsilon$. We have checked that this agrees with the scalings found numerically for the growth rate.

\section{Conclusions}
We have studied the collisionless tearing mode in the presence of significant shear flows, as parametrized by $\alpha = \delta u / \delta b$, where $\delta u$ and $\delta b$ are the amplitudes of the background velocity and magnetic field fluctuations, pointing in the $y$ (or L) direction and varying in the $x$ (or N) direction across the current sheet. We find that the growth rates depend strongly on $\alpha$, with a maximum growth rate given by (\ref{eq:gmax}),
\begin{equation}
    \frac{\gammatr a}{\vAy} \sim \frac{\rho_s^{1/n}d_e}{a^{1+1/n}}\frac{(1-\alpha^2)^{3/2-1/2n}}{\alpha^{2-2/n}}\frac{(1+\tau/Z)^{3/2n-1}}{(1+\alpha^2\tau/Z)^{1/n}},
\end{equation}
where $n=1$ (e.g. for a Harris-type profile) or $n=2$ (e.g. for a sinusoidal type profile). As $\alpha\to1$ (an exactly Alfv\'enic flow), $\gamma \to 0$. Moreover, for large ion-to-electron temperature ratios, $\tau=T_{0i}/T_{0e}$, the growth rate decreases with $\tau$, $\gammatr\propto \tau^{1/2n-1}$. With $n=1$ for simplicity, relative to the maximum tearing mode growth rate with $\alpha=0$, $\gamma_{0\mathrm{tr}}$ (see Eq.~\ref{eq:gmaxcless}), we have
\begin{equation}
    \frac{\gammatr}{\gamma_{0\mathrm{tr}}} = \frac{1-\alpha^2}{1+\alpha^2 \tau/Z}.
\end{equation}

\markup{As described in the introduction, a} remarkable absence of reconnection was reported in the near-Sun solar wind observed during Parker Solar Probe's first perihelion \citep{phan2020}. This has been confirmed in a comprehensive recent study of the PSP data by \cite{eriksson2024}, who found that reconnection was extremely rare more specifically in faster, high-ion-temperature solar wind emerging from coronal holes: this is also the wind that is typically highly Alfv\'enic \citep{damicis2015,damicis2021}, with highly correlated velocity and magnetic field fluctuations, $\delta \vu \sim \pm\delta \vb$ or $1-\alpha^2\ll1$ \citep{ervin2024}. Typically, this wind also has large $\tau=T_{0i}/T_{0e}$ \citep{shi2023}, probably due to higher ion heating in imbalanced turbulence due to the recently discovered helicity barrier \citep{meyrand2021,squire2022c}. Similar clustering was observed at larger heliocentric distances by \cite{fargette2023}, possibly with the same underlying cause.

We have shown that both these parameters typical of the Alfv\'enic wind, $1-\alpha^2\ll1$ and $\tau\gtrsim1$, suppress the tearing mode growth rate, and thus increase the reconnection onset time, comparable to $\gamma^{-1}$. We can roughly estimate a typical reconnection time from our expression for the growth rate: taking $n=1$ and typical values for the PSP Alfv\'enic wind, $\alpha\approx0.9$ and $\tau\approx 2$ \citep{chen2020,shi2021,shi2023,ervin2024}, we find $\gammatr/\gamma_{0\mathrm{tr}} \approx 0.07$. \markup{On the other hand, the eddy turnover time $\tau_{nl}\sim \lambda / \delta u(1-\alpha^2)$ in imbalanced turbulence is increased by the same factor $(1-\alpha^2)^{-1}$ \citep{schekochihin2022}, so if all the current sheets are produced by turbulence, the product $\gammatr\tau_{nl}$ is left relatively unchanged. If the development of current sheets is controlled by the eddy turnover time, then the overall prevalence of reconnection would not be suppressed by this mechanism. However, it is far from clear that this is the only timescale available. For example, sharp switchback boundaries may evolve instead due to the large-scale inhomogeneity in the system, on the solar wind expansion time of order $\tau_{exp} \sim R/U_{sw}$, where $R$ is the radial distance from the Sun and $U_{sw}$ is the solar wind velocity \citep{mallet2021}.}

\markup{Our work could be extended in several important ways to better describe the solar wind current sheets. First, KREHM cannot handle large equilibrium flows, and we are therefore limited to guide-field reconnection. Second, we have not allowed an equilibrium parallel flow shear, which is also often observed in the solar wind \citep{eriksson2024}. Third, KREHM is formally limited to small $\beta$. This is perhaps reasonable for the solar wind close to the Sun, but further out in the heliosphere $\beta$ is typically of order unity. Finally, we have also not allowed an equilibrium density gradient (again, often observed in the data), which can suppress reconnection \citep{swisdak2003} and also requires a more careful treatment of the matching between the ion and outer layers \citep{connor2019}. Nevertheless, our theory may help to explain the patchy suppression of reconnection in the PSP data, but further work is needed to generalize the equilibrium and to elucidate the formation timescale of the observed sheet-like structures.}

Declaration of Interests.-- The authors report no conflict of interest.

\markup{Acknowledgements.-- The authors would like to thank the anonymous reviewers for helpful comments that significantly improved the paper. AM, SE and MS were supported by NASA grant 80NSSC20K1284. JJ was funded by the U.S. Department of Energy under Contract No. DE-AC02-09CH1146 via an LDRD grant.}

\bibliographystyle{jpp}
\bibliography{mainbib2024}

\end{document}